\documentclass[10pt,aps,superscriptaddress,nofootinbib]{revtex4}
\usepackage{graphicx}
\usepackage{amsmath}
\usepackage{multirow}
\usepackage{float}
\usepackage{color}

\begin{document}

\begin{flushright}
  SMU-HEP-13-22\\
  Aug 22, 2013
\end{flushright}

\title{Differentiating the production mechanisms of the Higgs-like resonance
using inclusive observables at hadron colliders}

\author{Jun Gao}
\affiliation{Department of Physics, Southern Methodist University,
Dallas, TX 75275-0181, USA}
\email{jung@smu.edu}

\begin{abstract}
We present a study on differentiating direct production
mechanisms of the newly discovered Higgs-like boson at the LHC based
on several inclusive observables. The ratios introduced reveal the
parton constituents or initial state radiations involved in the
production mechanisms, and are directly sensitive to fractions of
contributions from different channels.
We select three benchmark models, including the SM Higgs boson, to
illustrate how the theoretical predictions of the above ratios are
different for the $gg$, $b\bar b(c\bar c)$, and $q\bar q$ (flavor
universal) initial states in the direct production. We study
implications of current Tevatron and LHC measurements. We also
show expectations from
further LHC measurements with high luminosities.

\noindent \textbf {Keywrords}: Higgs Physics, Beyond Standard Model \quad

\end{abstract}

\maketitle

\section{Introduction\label{sec:intro}}
Recently, a new resonance with a mass around 126 GeV has been
discovered by the ATLAS~\cite{Aad:2012tfa} and
CMS~\cite{Chatrchyan:2012ufa}. It is considered to be a highly
Standard Model (SM) Higgs-like particle with measured production
rate consistent with the SM Higgs boson through $\gamma\gamma$,
$ZZ^*$, $WW^*$, and $\tau\tau$
channels~\cite{Aad:2012tfa,Chatrchyan:2012ufa}. Although further
efforts are required in order to determine the features of the new
resonance, like the spin, couplings with SM particles, and
self-couplings. The spin-1 hypothesis is excluded by the observation
of the $\gamma\gamma$ decay mode according to the Landau-Yang
theorem~\cite{Landau:1948kw,Yang:1950rg}. Many proposals have been
suggested to distinguish between the spin-0 and spin-2 hypotheses
mainly focusing on the kinematic distributions, e.g, angular
distributions~\cite{Choi:2002jk,Gao:2010qx,DeRujula:2010ys,Englert:2010ud,Ellis:2012wg,Bolognesi:2012mm,Choi:2012yg,Ellis:2012jv,Englert:2012xt,Banerjee:2012ez,Modak:2013sb,Boer:2013fca,Frank:2013gca},
event shapes~\cite{Englert:2013opa} and other
observables~\cite{Boughezal:2012tz,Ellis:2012xd,Alves:2012fb,Geng:2012hy,Djouadi:2013yb}.
Recent measurements~\cite{ATLAS:2013xla,ATLAS:2013mla,Aad:2013xqa,CMS:xwa} show a favor
of spin-0 over specific spin-2 scenarios. As for the couplings, the
current direct information or constraints are for the relative
strength between different observed channels, i.e., $\gamma\gamma$,
$ZZ^*$, $WW^*$, and
$\tau\tau$~\cite{Aad:2012tfa,Chatrchyan:2012ufa}. Without knowing
the total decay width and rates from other unobserved channels it is
difficult to determine the absolute strength of the couplings of the
new resonance at the LHC. Or later we can further measure the
couplings through a combined analysis after the observation of the
associated production modes with the SM $W$ and $Z$ bosons or the
vector-boson fusion (VBF) production
mode~\cite{Plehn:2001nj,Giardino:2012ww,Rauch:2012wa,Azatov:2012rd,Low:2012rj,Carmi:2012in,Plehn:2012iz,Djouadi:2012rh}.

Among all the couplings of the new resonance, the ones with gluons or quarks
are important but difficult to be measured at the LHC since the corresponding decay
modes consist of two jets, which suffer from huge QCD backgrounds at the LHC
even for the heavy-quark (charm or bottom quark) jets. Moreover, it is extremely
hard to discriminate the couplings with gluons and light-quarks from the
resonance decay. This relates to the answer to a more essential question, i.e.,
the direct production of the new resonance is dominated by the gluon fusion
or quark annihilation. In the SM, the loop-induced gluon fusion is dominant
while the heavy-quark annihilation only contributes at a percent level. As
for other hypotheses, like in the two Higgs doublet models, the heavy-quark
contributions can be largely enhanced~\cite{Djouadi:2005gj,Meng:2012uj}, or
in the graviton-like cases~\cite{ArkaniHamed:1998rs,Randall:1999ee}, the light-quark
contributions are important as well.

Similar as in the determination of the spin of the new resonance, we
can use the angular distributions of the observed decay products,
like $\gamma\gamma$, $ZZ^*$, and $WW^*$, to differentiate the $gg$
and $q\bar q$ production mechanisms as
in~\cite{ATLAS:2013xla,ATLAS:2013mla,Aad:2013xqa,CMS:xwa}. But the analyses are highly
model-dependent, i.e., the angular distributions are sensitive to
the spin of the resonance as well as the structures of the couplings
with the decay products~\cite{Gao:2010qx}. On another hand, since
these two production mechanisms depend on different flavor
constituents of the parton distribution functions (PDFs), they may
show distinguishable behaviors by looking at the ratios of the event
rate at different colliders or center-of-mass energies as previously shown
in~\cite{Mangano:2012mh} for various SM processes at the LHC including for
the SM Higgs boson, or
the rapidity distribution of the resonance. Even more ambitious, we
may look at the production of the resonance in association with an
additional photon or jet from initial state radiations which are
presumably to be different for the gluon and quark initial states.
Unlike the case of the angular distributions, all these
observables are insensitive to details of couplings with the decay products.
Thus they may serve as good
discriminators of the direct production mechanisms of the new
resonance.

Based on the above ideas we present a study of using inclusive observables to
discriminate the mechanisms of the direct production of the new resonance, including
both the theoretical predictions and the experimental feasibilities.
In Section~\ref{sec:set}, we describe the benchmark models of the production mechanisms
studied in this paper, and introduce several inclusive observables that are used
in our study. Section~\ref{sec:ben} compares the theoretical predictions of the observables
from different models. In Section~\ref{sec:exp} we discuss the applications on current
experimental data from the Tevatron and LHC, and also future measurements at the LHC.
Section~\ref{sec:con} is a brief conclusion.

\section{Model setups and inclusive observables\label{sec:set}}

We select three benchmark models in the study, including the pure SM case, an
alternative spin-0 resonance with enhanced couplings to the charm and
bottom quarks, and a spin-2 resonance with universal couplings to all the quarks.
As explained in the introduction, our analyses mainly rely on the fractions of
the $gg$ and $q\bar q$ contributions in the production mechanism and are insensitive
to details of couplings with the decay products. More precisely, the relevant effective
couplings for the spin-0 cases are given by
\begin{equation}
{\mathcal L}_{spin-0}={g_1^{(0)}\over v}HG^{\mu\nu}G_{\mu\nu}+{g_2^{(0)}\over v}
(m_cH\bar{\Psi}_c\Psi_c+m_bH\bar{\Psi}_b\Psi_b),
\end{equation}
and for the spin-2 case by~\cite{Englert:2012xt}
\begin{equation}
{\mathcal L}_{spin-2}=g_1^{(2)}Y_{\mu\nu}T_{G}^{\mu\nu}+g_2^{(2)}Y_{\mu\nu}T_{q}
^{\mu\nu},
\end{equation}
with $H$ being the scalar particle, $Y_{\mu\nu}$ the general spin-2
fields~\cite{spin2a,spin2b}, $G_{\mu\nu}$ the field strength of QCD,
and $\Psi_{c,b}$ the charm and bottom quarks. We choose
graviton-inspired couplings for the spin-2 case with
$T_{G}^{\mu\nu}$ and $T_{q}^{\mu\nu}$ being the energy-momentum
tensors of the gluon and quarks (flavor universal) as can be found
in~\cite{Hagiwara:2008jb}. Here we suppress all other couplings of
the resonance with the $W$, $Z$ bosons, photon, and $\tau$ lepton,
which are adjusted to satisfy the corresponding decay branching
ratios observed~\cite{Aad:2012tfa,Chatrchyan:2012ufa}, especially
the couplings with photons should be suppressed in order to be
consistent with the experimental measurements. We work under an
effective Lagrangian approach and will not discuss about the
possible UV completion of the theory.

For model A, the pure
SM, we have
\begin{equation}
v=246\, {\rm GeV},\,\, g_1^{(0)}={\alpha_s\over 12\pi},\,\, g_2^{(0)}=1,\,\,
m_{c(b)}=0.634(2.79)\, {\rm GeV},
\end{equation}
where $g_1^{(0)}$ are evaluated at the LO in the infinite top quark
mass limit, and the heavy-quark masses are $\overline{\rm MS}$
running mass at the resonance mass $m_X=126\,{\rm
GeV}$~\cite{Beringer:1900zz,Chetyrkin:2000yt}. From a
phenomenological point of view, we introduce model B, the
heavy-quark dominant case with $g_1^{(0)}=0$. Note that $g_1^{(0)}$
always receives non-zero contributions from the heavy-quark loops
proportional to $g_2^{(0)}$. However, in global analyses of the
Higgs couplings~\cite{Carmi:2012in,Djouadi:2012rh}, it is always
treated as another free parameter that could in principle vanish,
since its actual value depends on details of the underlying new
physics. Thus model B is a phenomenological simplification of models
with heavy-quark annihilation dominant in the production, e.g.,
supersymmetric models with large $\tan \beta$~\cite{Djouadi:2005gj}.
The absolute value of $g_2^{(0)}$ is irrelevant for the study here.
Similar for model C, the spin-2 case, we set $g_1^{(2)}=0$ with the
production dominated by the light quarks. It is shown that a spin-2
model with minimal couplings~\cite{Gao:2010qx} to the vector bosons
has been ruled out by both the ATLAS and CMS despite of the
production mechanism~\cite{Aad:2013xqa,CMS:xwa}. The measurements
utilize angular distributions of final states from decay vector
bosons. As shown in~\cite{Gao:2010qx}, these angular distributions
are sensitive to detailed structures of the couplings to the vector
bosons. Thus the exclusion could not be applied to a general
spin-2 model involving much more free parameters in the vector boson
couplings~\cite{Gao:2010qx}. In contrast the observables
introduced below are independent of the couplings to the decay vector
bosons.

The inclusive observables we studied can be divided into three
categories. First one is the ratio of the inclusive cross sections
of the direct production, $R^{1}$, including the cross sections at
the Tevatron, and at the LHC with different center-of-mass energies.
The second one is the ratio of the direct production cross sections
in the inner and full rapidity region of the produced resonance,
$R^{2}$. These two observables probe the production mechanisms
through the differences of the relevant PDFs. The third observable,
$R^{3}$, is the ratio of the production cross section of the
resonance in association with a photon to the one of the direct
production. It differentiates the production channels by measuring
the initial state radiations. For the calculation of $R^{3}$ we
neglect the small explicit couplings of the new resonance with
photons in the production. Other observables that might be sensitive
to the production mechanisms are related to the initial state QCD
radiations, like the $p_T$ spectrum or jet-bin cross
sections~\cite{Dittmaier:2012vm,Wiesemann:2012ij} of the resonance,
which are again different for the $gg$ and $q\bar q$ initial states.
But that will be even more challenging in both the theory
predictions and experimental measurements.

\section{Benchmark comparisons\label{sec:ben}}

\subsection{Ratios of the total cross section\label{sec:ben1}}
Here we calculate the total cross sections of the direct production of the new
resonance at
the Tevatron and LHC with $\sqrt s=7$, 8, and 14 TeV. At the leading order (LO),
they are related to the following parton-parton luminosities,
\begin{align}\label{eq:lum}
&L_{gg}(\tau)=\int_{\tau}^1\frac{dx_1}{x_1}\int_{\tau/x_1}^1\frac{dx_2}{x_2}\tau^2
f_{g/h_1}(x_1, \mu_f)f_{g/h_2}(x_2, \mu_f)\delta(x_1x_2-\tau), \nonumber \\
&L_{c\bar c(b \bar b)}(\tau)=\int_{\tau}^1\frac{dx_1}{x_1}\int_{\tau/x_1}^1\frac{dx_2}{x_2}\tau^2
[f_{c(b)/h_1}(x_1, \mu_f)f_{\bar c(\bar b)/h_2}(x_2, \mu_f)+h_1\leftrightarrow h_2]\delta(x_1x_2-\tau),
\nonumber\\
&L_{q\bar q}(\tau)=\sum_{q}\int^1_{\tau} \frac{dx_1}{x_1} \int^1_{\tau/x_1} \frac{dx_2}{x_2} \tau^2
[f_{q/h_1}(x_1, \mu_f)f_{\bar q/h_2}(x_2, \mu_f)+h_1\leftrightarrow h_2]\delta(x_1x_2-\tau),
\end{align}
where $\tau=m_X^2/s$, $x_{1,2}$ are the momentum fractions. $\mu_f$
is the factorization scale and set to $m_X$ in our calculations.
$f_{i/h}(x)$ are the PDFs, and the sum in $L_{q\bar q}$ runs over
all the 5 active quark flavors. Thus the typical Bjorken $x\sim
m_X/\sqrt s$ are about 0.06, 0.018, 0.016, and 0.009 at the
Tevatron, LHC 7, 8, and 14 TeV. While beyond LO, there are also
contributions from other flavor combinations subject to different
$x_1-x_2$ constraints. We select 5 ratios from all the cross
sections, $R^1_{L7/T}=\sigma({\rm LHC}\,7\, {\rm TeV})/\sigma({\rm
Tevatron})$, similar for $R^1_{L8/T}$, $R^1_{L14/T}$,
$R^1_{L14/L7}$, and $R^1_{L14/L8}$. The cross sections for models A
and B can be calculated up to next-to-next-to-leading order (NNLO)
in QCD using the numerical code iHixs1.3~\cite{Anastasiou:2011pi}.
While it is only calculated at the LO for the model C. Note that the
ratios $R^1$ at the LO are totally determined by the behaviors of
the parton-parton luminosities in Eq.~(\ref{eq:lum}) and are
independent of the detailed structures of the couplings, while at
higher orders they may show slight dependence on the couplings. We
set the renormalization scale to $m_X=126\,{\rm GeV}$ as well, and
use the most recent NNLO PDFs including CT10~\cite{Gao:2013xoa},
MSTW 2008~\cite{Martin:2009iq}, and NNPDF2.3~\cite{Ball:2012cx}. The
PDF and $\alpha_s$ uncertainties are calculated and combined using
the prescription in~\cite{Ball:2012wy}.

In Table.~\ref{tab:r1a} we show the predicted ratios $R^1$ for the
SM Higgs boson from different PDF groups. It can be seen that the
current uptodate NNLO PDFs give pretty close results for the ratios.
The combined PDF+$\alpha_s$ uncertainties are about 7\% for the
ratios of the NNLO cross sections at the LHC over Tevatron due to
the relatively large uncertainties of the gluon PDF at the large $x$
region. While the uncertainties are reduced to a level of about 2\%
for the ratios at the LHC. Theoretical uncertainties due to the
missing higher order QCD corrections can be estimated by looking at
the differences of the results at different orders, which are
smaller compared to the combined PDF+$\alpha_s$ uncertainties and
are not considered in our analysis. Tables.~\ref{tab:r1b} and
\ref{tab:r1c} show similar results for the model B and C. The heavy
quark PDFs are mostly generated through the evolution of the gluon
PDF. Thus the results of the model B are close to the SM case. The
model C predicts very different results compared to the SM or model
B, for the ratios of the NNLO cross sections at the LHC over
Tevatron, and also shows smaller uncertainties, since the cross
sections are dominated by the light quark scattering.

\begin{table}[h!]
  \begin{center}
      \begin{tabular}{c|ccc|ccc|ccc|c}
        \hline \hline
        \multirow{2}{*}{model A}
    &\multicolumn{3}{c}{CT10}&\multicolumn{3}{c}{MSTW08}&\multicolumn{3}{c}{NNPDF2.3}&Combined \\
        \cline{2-11}
          & LO & NLO & NNLO & LO & NLO & NNLO & LO &NLO &NNLO& NNLO \\
        \hline
        $R^1_{L7/T}$ & $17.9^{+0.8}_{-1.0}$& $17.5^{+0.8}_{-0.9}$ & $17.0^{+0.7}_{-0.9}$ & $18.1^{+0.5}_{-0.5}$ & $17.7^{+0.5}_{-0.5}$ & $17.2^{+0.5}_{-0.5}$ & $18.6^{+0.6}_{-0.6}$&$18.1^{+0.6}_{-0.5}$ & $17.5^{+0.5}_{-0.5}$ & $17.1^{+1.1}_{-1.1}$\\
        \hline
        $R^1_{L8/T}$ & $22.9^{+1.1}_{-1.3}$& $22.4^{+1.0}_{-1.2}$ & $21.7^{+1.0}_{-1.2}$ & $23.2^{+0.7}_{-0.7}$& $22.6^{+0.7}_{-0.7}$ &$21.9^{+0.7}_{-0.7}$ & $23.9^{+0.8}_{-0.8}$& $23.2^{+0.7}_{-0.7}$ & $22.4^{+0.7}_{-0.7}$ &$21.8^{+1.5}_{-1.5}$ \\
        \hline
        $R^1_{L14/T}$ & $59.9^{+3.4}_{-4.1}$& $58.5^{+3.1}_{-3.8}$ & $56.3^{+3.0}_{-3.6}$ & $60.7^{+2.3}_{-2.2}$& $59.3^{+2.2}_{-2.1}$ & $57.0^{+2.1}_{-2.0}$ & $62.2^{+2.4}_{-2.3}$ & $60.6^{+2.2}_{-2.1}$ & $58.1^{+2.1}_{-2.0}$ &$56.6^{+4.3}_{-4.3}$ \\
        \hline
        $R^1_{L14/L7}$ & $3.34^{+0.04}_{-0.05}$& $3.35^{+0.04}_{-0.05}$ & $3.32^{+0.04}_{-0.05}$ & $3.35^{+0.03}_{-0.03}$& $3.35^{+0.03}_{-0.03}$ & $3.32^{+0.03}_{-0.03}$ & $3.34^{+0.03}_{-0.03}$& $3.34^{+0.03}_{-0.02}$ & $3.31^{+0.02}_{-0.02}$ & $3.31^{+0.05}_{-0.05}$\\
        \hline
        $R^1_{L14/L8}$ & $2.61^{+0.02}_{-0.03}$& $2.62^{+0.02}_{-0.03}$ & $2.60^{+0.02}_{-0.03}$ & $2.62^{+0.02}_{-0.02}$& $2.62^{+0.02}_{-0.02}$ & $2.60^{+0.02}_{-0.02}$ & $2.61^{+0.02}_{-0.02}$& $2.61^{+0.02}_{-0.01}$ & $2.59^{+0.01}_{-0.01}$ & $2.59^{+0.03}_{-0.03}$\\
        \hline \hline
      \end{tabular}
  \end{center}
  \vspace{-3ex}
  \caption{\label{tab:r1a}Predicted ratios $R^1$ at different orders
  from various PDFs with the PDF+$\alpha_s$ uncertainties at 68\% C.L.
  for the case of pure SM.}
\end{table}

\begin{table}[h!]
  \begin{center}
      \begin{tabular}{c|ccc|ccc|ccc|c}
        \hline \hline
        \multirow{2}{*}{model B}
    &\multicolumn{3}{c}{CT10}&\multicolumn{3}{c}{MSTW08}&\multicolumn{3}{c}{NNPDF2.3}&Combined \\
        \cline{2-11}
          & LO & NLO & NNLO & LO & NLO & NNLO & LO &NLO &NNLO& NNLO \\
        \hline
        $R^1_{L7/T}$ & $23.0^{+1.5}_{-1.8}$& $22.7^{+1.6}_{-1.9}$ & $23.4^{+1.7}_{-2.0}$ & $23.5^{+1.0}_{-1.0}$ & $23.2^{+1.1}_{-1.1}$ & $24.0^{+1.2}_{-1.2}$ & $24.6^{+1.2}_{-1.2}$&$24.4^{+1.3}_{-1.2}$ & $25.3^{+1.5}_{-1.4}$ & $24.2^{+3.2}_{-3.2}$\\
        \hline
        $R^1_{L8/T}$ & $29.8^{+2.1}_{-2.5}$& $29.4^{+2.2}_{-2.6}$ & $30.4^{+2.4}_{-2.8}$ & $30.5^{+1.4}_{-1.4}$& $30.0^{+1.5}_{-1.5}$ &$31.2^{+1.7}_{-1.7}$ & $32.0^{+1.7}_{-1.6}$& $31.6^{+1.8}_{-1.7}$ & $33.0^{+2.0}_{-1.9}$ &$31.4^{+4.3}_{-4.3}$ \\
        \hline
        $R^1_{L14/T}$ & $81.2^{+6.6}_{-7.8}$& $79.4^{+6.8}_{-7.9}$ & $82.8^{+7.5}_{-8.6}$ & $83.1^{+4.7}_{-4.6}$& $81.6^{+5.0}_{-4.8}$ & $85.3^{+5.6}_{-5.4}$ & $87.4^{+5.2}_{-4.9}$ & $85.8^{+5.5}_{-5.1}$ & $90.0^{+6.2}_{-5.7}$ &$85.6^{+13.1}_{-13.1}$ \\
        \hline
        $R^1_{L14/L7}$ & $3.53^{+0.06}_{-0.07}$& $3.50^{+0.06}_{-0.07}$ & $3.54^{+0.06}_{-0.08}$ & $3.54^{+0.04}_{-0.04}$& $3.52^{+0.04}_{-0.04}$ & $3.55^{+0.05}_{-0.04}$ & $3.54^{+0.04}_{-0.04}$& $3.52^{+0.04}_{-0.04}$ & $3.55^{+0.04}_{-0.04}$ & $3.53^{+0.09}_{-0.09}$\\
        \hline
        $R^1_{L14/L8}$ & $2.72^{+0.03}_{-0.04}$& $2.70^{+0.04}_{-0.04}$ & $2.72^{+0.04}_{-0.04}$ & $2.73^{+0.02}_{-0.02}$& $2.71^{+0.03}_{-0.02}$ & $2.73^{+0.03}_{-0.03}$ & $2.73^{+0.02}_{-0.02}$& $2.71^{+0.02}_{-0.02}$ & $2.73^{+0.02}_{-0.02}$ & $2.72^{+0.05}_{-0.05}$\\
        \hline \hline
      \end{tabular}
  \end{center}
  \vspace{-3ex}
  \caption{\label{tab:r1b}Predicted ratios $R^1$ at different orders
  from various PDFs with the PDF+$\alpha_s$ uncertainties at 68\% C.L.
  for model B.}
\end{table}

\begin{table}[h!]
  \begin{center}
      \begin{tabular}{c|c|c|c|c}
        \hline \hline
        \multirow{2}{*}{model C}
    &\multicolumn{1}{c}{CT10}&\multicolumn{1}{c}{MSTW08}&\multicolumn{1}{c}{NNPDF2.3}&Combined \\
        \cline{2-5}
          & LO & LO  & LO & LO \\
        \hline
        $R^1_{L7/T}$ & $3.96^{+0.07}_{-0.06}$& $4.00^{+0.04}_{-0.06}$  & $3.95^{+0.06}_{-0.05}$ & $3.98^{+0.10}_{-0.10}$\\
        \hline
        $R^1_{L8/T}$ & $4.68^{+0.08}_{-0.08}$& $4.72^{+0.05}_{-0.07}$ & $4.67^{+0.07}_{-0.06}$ &$4.70^{+0.12}_{-0.12}$ \\
        \hline
        $R^1_{L14/T}$ & $9.17^{+0.20}_{-0.20}$ & $9.19^{+0.13}_{-0.16}$ & $9.10^{+0.14}_{-0.12}$  &$9.18^{+0.25}_{-0.25}$ \\
        \hline
        $R^1_{L14/L7}$ & $2.32^{+0.02}_{-0.02}$ & $2.30^{+0.01}_{-0.01}$ & $2.30^{+0.01}_{-0.01}$ & $2.31^{+0.02}_{-0.02}$\\
        \hline
        $R^1_{L14/L8}$ & $1.96^{+0.01}_{-0.01}$ & $1.94^{+0.01}_{-0.01}$ & $1.95^{+0.01}_{-0.01}$ & $1.96^{+0.02}_{-0.02}$\\
        \hline \hline
      \end{tabular}
  \end{center}
  \vspace{-3ex}
  \caption{\label{tab:r1c}Predicted ratios $R^1$ at the LO
  from various PDFs with the PDF+$\alpha_s$ uncertainties at 68\% C.L.
  for model C.}
\end{table}

\subsection{Centrality ratio\label{sec:ben2}}
\begin{figure}[h]
  \begin{center}
    \includegraphics[width=0.8\textwidth]{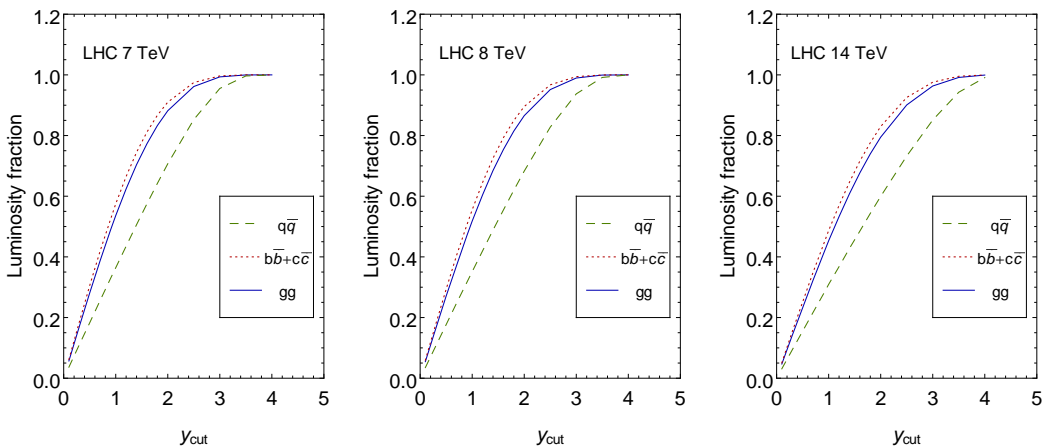}
  \end{center}
  \vspace{-5ex}
  \caption{\label{fig:r2}
  Luminosity fractions as a function of the rapidity cutoff at the LHC with
  different center-of-mass energies.}
\end{figure}

At the LO, the rapidity of the produced resonance in the lab frame is given by,
$y=\ln(x_1/x_2)/2$, or equivalently $y=\ln((1+\beta)/(1-\beta))/2$, where $\beta$
is the boost of the resonance. We define the centrality $R^2$ as the
ratio of the production cross section in the central region (with $|y|<1$)
to the one in the full rapidity region, which are related to the corresponding
ratio of the parton-parton luminosities at the LO, $L(\tau, |y|<1)/L(\tau)$.
For illustration purpose, we show the above luminosity ratio as functions
of the rapidity cutoff in Fig.~\ref{fig:r2} for different parton combinations shown in Eq.~(\ref{eq:lum}).

The calculated centrality ratios for the models A, B, and C are listed in Tables.~\ref{tab:r2a}-\ref{tab:r2c}
for different PDFs at the LHC. Again for the pure SM case, the predictions are
at the NNLO in QCD from HNNLO1.3 code~\cite{Catani:2007vq}. Others are only calculated at the LO.
Here we simply choose the central region of $|y|<1$ for the definition of $R^2$.
In principle one can find the optimized value that gives largest distinctions of the
three models. Similar to the case of $R^1$, the models A and B give close results of $R^2$ but with
larger uncertainties compared to $R^1$. The differences of the predictions from the model C with the
ones from the model A or B are still significant.

\begin{table}[h!]
  \begin{center}
      \begin{tabular}{c|ccc|ccc|ccc|c}
        \hline \hline
        \multirow{2}{*}{model A}
    &\multicolumn{3}{c}{CT10}&\multicolumn{3}{c}{MSTW08}&\multicolumn{3}{c}{NNPDF2.3}&Combined \\
        \cline{2-11}
          & LO & NLO & NNLO & LO & NLO & NNLO & LO &NLO &NNLO& NNLO \\
        \hline
        $R^2_{L7}$ & $0.536^{+0.009}_{-0.013}$& $0.536^{+0.009}_{-0.013}$ & $0.533^{+0.009}_{-0.013}$ & $0.538^{+0.005}_{-0.007}$ &$0.537^{+0.005}_{-0.007}$  & $0.538^{+0.005}_{-0.007}$ & $0.548^{+0.008}_{-0.008}$&$0.546^{+0.008}_{-0.008}$ & $0.547^{+0.008}_{-0.008}$ & $0.539^{+0.018}_{-0.018}$\\
        \hline
        $R^2_{L8}$ & $0.518^{+0.009}_{-0.012}$& $0.519^{+0.009}_{-0.012}$ & $0.526^{+0.009}_{-0.012}$ & $0.518^{+0.009}_{-0.003}$& $0.522^{+0.009}_{-0.003}$ & $0.532^{+0.009}_{-0.003}$ & $0.529^{+0.008}_{-0.008}$& $0.530^{+0.008}_{-0.008}$ & $0.538^{+0.008}_{-0.008}$ &$0.531^{+0.017}_{-0.017}$ \\
        \hline
        $R^2_{L14}$ & $0.453^{+0.007}_{-0.008}$& $0.453^{+0.007}_{-0.008}$  & $0.450^{+0.007}_{-0.008}$ & $0.454^{+0.004}_{-0.004}$& $0.454^{+0.004}_{-0.004}$ & $0.452^{+0.004}_{-0.004}$ & $0.461^{+0.005}_{-0.005}$ & $0.460^{+0.005}_{-0.005}$  & $0.458^{+0.005}_{-0.005}$  &$0.453^{+0.012}_{-0.012}$ \\
        \hline \hline
      \end{tabular}
  \end{center}
  \vspace{-3ex}
  \caption{\label{tab:r2a}Predicted ratios $R^2$ at different orders
  from various PDFs with the PDF+$\alpha_s$ uncertainties at 68\% C.L.
  for the case of pure SM.}
\end{table}

\begin{table}[h!]
  \begin{center}
      \begin{tabular}{c|c|c|c|c}
        \hline \hline
        \multirow{2}{*}{model B}
    &\multicolumn{1}{c}{CT10}&\multicolumn{1}{c}{MSTW08}&\multicolumn{1}{c}{NNPDF2.3}&Combined \\
        \cline{2-5}
          & LO & LO  & LO & LO \\
        \hline
        $R^2_{L7}$ & $0.575^{+0.012}_{-0.017}$& $0.578^{+0.006}_{-0.008}$  & $0.592^{+0.010}_{-0.010}$ & $0.580^{+0.023}_{-0.023}$\\
        \hline
        $R^2_{L8}$ & $0.555^{+0.012}_{-0.015}$& $0.556^{+0.014}_{-0.004}$ & $0.571^{+0.010}_{-0.010}$ &$0.561^{+0.022}_{-0.022}$ \\
        \hline
        $R^2_{L14}$ & $0.487^{+0.009}_{-0.011}$ & $0.489^{+0.005}_{-0.006}$ & $0.498^{+0.007}_{-0.007}$  &$0.490^{+0.016}_{-0.016}$ \\
        \hline \hline
      \end{tabular}
  \end{center}
  \vspace{-3ex}
  \caption{\label{tab:r2b}Predicted ratios $R^2$ at the LO
  from various PDFs with the PDF+$\alpha_s$ uncertainties at 68\% C.L.
  for model B.}
\end{table}

\begin{table}[h!]
  \begin{center}
      \begin{tabular}{c|c|c|c|c}
        \hline \hline
        \multirow{2}{*}{model C}
    &\multicolumn{1}{c}{CT10}&\multicolumn{1}{c}{MSTW08}&\multicolumn{1}{c}{NNPDF2.3}&Combined \\
        \cline{2-5}
          & LO & LO  & LO & LO \\
        \hline
        $R^2_{L7}$ & $0.364^{+0.004}_{-0.005}$& $0.358^{+0.005}_{-0.002}$  & $0.361^{+0.002}_{-0.002}$ & $0.362^{+0.007}_{-0.007}$\\
        \hline
        $R^2_{L8}$ & $0.351^{+0.004}_{-0.005}$& $0.345^{+0.002}_{-0.003}$ & $0.348^{+0.002}_{-0.002}$ &$0.349^{+0.008}_{-0.008}$ \\
        \hline
        $R^2_{L14}$ & $0.309^{+0.004}_{-0.005}$ & $0.303^{+0.002}_{-0.003}$ & $0.309^{+0.002}_{-0.002}$  &$0.306^{+0.007}_{-0.007}$ \\
        \hline \hline
      \end{tabular}
  \end{center}
  \vspace{-3ex}
  \caption{\label{tab:r2c}Predicted ratios $R^2$ at the LO
  from various PDFs with the PDF+$\alpha_s$ uncertainties at 68\% C.L.
  for model C.}
\end{table}

\subsection{Associated production\label{sec:ben3}}
\begin{figure}[h]
  \begin{center}
    \includegraphics[width=0.8\textwidth]{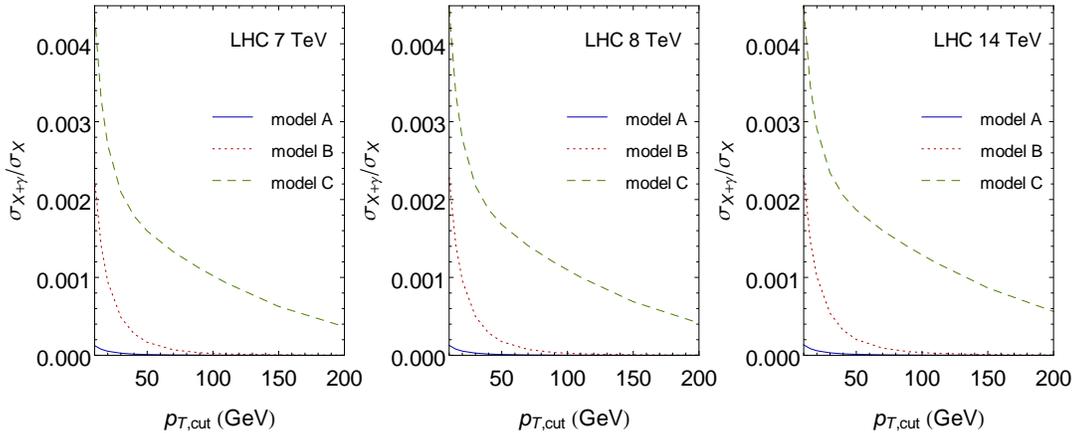}
  \end{center}
  \vspace{-5ex}
  \caption{\label{fig:r3}
  Ratios of the cross sections of the associated production to the ones of the direct production
  as functions of the $p_T$ cut of the photon, at the LHC with different
  center-of-mass energies.}
\end{figure}

Here we consider the ratios of the cross sections for the resonance production
in association with a photon to the ones of the direct production, $R^3\equiv\sigma_{X+\gamma}/\sigma_{X}$. The advantage
is that for the case of the SM, this associated production mode is largely suppressed
with main contributions from the $b\bar b$ annihilation at the LHC~\cite{Abbasabadi:1997zr}. While for
models B and C, the associated production is only suppressed by the QED couplings
even though the statistics are low at the LHC.
The calculations for the associated production are performed at the LO. Thus, for consistency
we use the LO cross sections of the direct production as well. Moreover, for
the model C, we apply a form factor~\cite{Frank:2013gca}
\begin{equation}
F=\left(\frac{\Lambda^2}{\hat s+\Lambda^2}\right)^5,
\end{equation}
to the associated production by multiplying it with the squared amplitudes since
the effective operator there violates unitarity above a certain energy scale. Here
$\hat s$ is the square of the partonic center-of-mass energy, and we choose the
cutoff scale $\Lambda$ to be $800\,{\rm GeV}$.
We select the events from the associated production with a rapidity cut of
$|y_{\gamma}|<2$ and a transverse momentum cut $p_{T,\gamma}>15\, {\rm GeV}$
on the photon. Here we adopt a relatively lower $p_T$ cut on the photon in order
to maximize the statistics of the associated production. Fig.~\ref{fig:r3} shows
ratios of the cross sections of associated production to the ones of the direct production
as functions of the $p_T$ cut of the photon at the LHC with different
center-of-mass energies. It can be seen that for the SM case, the cross sections
of the associated production are negligible, less then $10^{-4}$ times the
cross sections of the direct production. While for models B and C the ratios are
larger by an order of magnitude comparing to the SM, and the associated production
may be observable at the LHC. For lower $p_T$ cutoff the ratios from models B and C
are close. At moderate or high $p_T$ cutoff the ratios from model C are larger
due to the power enhancement from high dimension operators, and are sensitive to the
form factor applied and the UV completion of the theory. The central values and the PDF+$\alpha_s$ uncertainties of
$R^3$ predicted in different models are listed in Tables.~\ref{tab:r3a}-\ref{tab:r3c}.

We may also utilize production of the resonance in association with
a jet, and study the effects on observables like jet-bin (jet-veto)
fractions, $p_T$ distribution of the resonance as recently measured
in~\cite{TheATLAScollaboration:2013eia}. The cross sections of
associated production with a jet are much larger compared to the
case of a photon due to the strong couplings as well as opening of
new partonic channels. Especially for the SM case, $gg$ channel now
contributes and dominates over all others. Similarly, we consider a
ratio of the one-jet inclusive cross sections to the total inclusive
ones, $\sigma_{X+jet}/\sigma_{X}$. For example, using both LO cross
sections, we obtain the ratio as 0.355 (0.128) for model A (B) at
the LHC 8 TeV. Here we require a jet to have $|y|<2$ and $p_T>30\,
{\rm GeV}$. We can see the ratio is larger for the SM case, in
contrary with the case of a photon, because of the stronger
radiations from gluon initial states and the high dimension
effective operators. Thus this ratio may have some discrimination
powers on different production mechanisms. At the same time it also
has larger theoretical and experimental uncertainties associated
with the jet. The resummed $p_T$ spectrums of the resonance produced
through $gg$ and $b\bar b$ initial states have been predicted
in~\cite{Glosser:2002gm,Bozzi:2003jy,Field:2004tt} and
\cite{Field:2004nc,Belyaev:2005bs} respectively. Shapes of the two
distributions are very similar with both peak located around $10\sim
20$ GeV at the LHC for a resonance mass of about 120 GeV. Note that
experimentally the jet may fake a photon with a rate depending on
both the kinematics and photon isolation criteria. For the SM case,
this may induce non-negligible contributions to the photon
associated production. We will not discuss these possibilities in
the analysis since they are highly dependent on details of the
experiments.

\begin{table}[h!]
  \begin{center}
      \begin{tabular}{c|c|c|c|c}
        \hline \hline
        \multirow{1}{*}{model A }
    &\multicolumn{1}{c}{CT10}&\multicolumn{1}{c}{MSTW08}&\multicolumn{1}{c}{NNPDF2.3}&Combined \\
        \cline{2-5}
         $\times10^{-3}$ & LO & LO  & LO & LO \\
        \hline
        $R^3_{L7}$ & $0.077^{+0.003}_{-0.003}$& $0.075^{+0.002}_{-0.002}$  & $0.077^{+0.002}_{-0.002}$ & $0.077^{+0.004}_{-0.004}$\\
        \hline
        $R^3_{L8}$ & $0.079^{+0.003}_{-0.002}$& $0.077^{+0.002}_{-0.002}$ & $0.080^{+0.002}_{-0.002}$ &$0.079^{+0.004}_{-0.004}$ \\
        \hline
        $R^3_{L14}$ & $0.085^{+0.002}_{-0.002}$ & $0.083^{+0.002}_{-0.002}$ & $0.086^{+0.002}_{-0.002}$  &$0.085^{+0.004}_{-0.004}$ \\
        \hline \hline
      \end{tabular}
  \end{center}
  \vspace{-3ex}
  \caption{\label{tab:r3a}Predicted ratios $R^3$ at the LO
  from various PDFs with the PDF+$\alpha_s$ uncertainties at 68\% C.L.
  for the case of pure SM.}
\end{table}

\begin{table}[h!]
  \begin{center}
      \begin{tabular}{c|c|c|c|c}
        \hline \hline
        \multirow{1}{*}{model B }
    &\multicolumn{1}{c}{CT10}&\multicolumn{1}{c}{MSTW08}&\multicolumn{1}{c}{NNPDF2.3}&Combined \\
        \cline{2-5}
         $\times10^{-3}$ & LO & LO  & LO & LO \\
        \hline
        $R^3_{L7}$ & $1.407^{+0.014}_{-0.014}$& $1.398^{+0.008}_{-0.006}$  & $1.405^{+0.007}_{-0.007}$ & $1.408^{+0.017}_{-0.017}$\\
        \hline
        $R^3_{L8}$ & $1.424^{+0.014}_{-0.013}$& $1.417^{+0.007}_{-0.006}$ & $1.426^{+0.007}_{-0.008}$ &$1.425^{+0.016}_{-0.016}$ \\
        \hline
        $R^3_{L14}$ & $1.467^{+0.013}_{-0.015}$ & $1.464^{+0.003}_{-0.007}$ & $1.478^{+0.008}_{-0.008}$  &$1.470^{+0.017}_{-0.017}$ \\
        \hline \hline
      \end{tabular}
  \end{center}
  \vspace{-3ex}
  \caption{\label{tab:r3b}Predicted ratios $R^3$ at the LO
  from various PDFs with the PDF+$\alpha_s$ uncertainties at 68\% C.L.
  for model B.}
\end{table}

\begin{table}[h!]
  \begin{center}
      \begin{tabular}{c|c|c|c|c}
        \hline \hline
        \multirow{1}{*}{model C }
    &\multicolumn{1}{c}{CT10}&\multicolumn{1}{c}{MSTW08}&\multicolumn{1}{c}{NNPDF2.3}&Combined \\
        \cline{2-5}
         $\times10^{-3}$ & LO & LO  & LO & LO \\
        \hline
        $R^3_{L7}$ & $3.291^{+0.057}_{-0.066}$& $3.365^{+0.030}_{-0.019}$  & $3.344^{+0.034}_{-0.036}$ & $3.302^{+0.089}_{-0.084}$\\
        \hline
        $R^3_{L8}$ & $3.364^{+0.058}_{-0.066}$& $3.438^{+0.025}_{-0.023}$ & $3.420^{+0.035}_{-0.037}$ &$3.376^{+0.085}_{-0.085}$ \\
        \hline
        $R^3_{L14}$ & $3.458^{+0.057}_{-0.061}$ & $3.523^{+0.026}_{-0.022}$ & $3.521^{+0.037}_{-0.039}$  &$3.474^{+0.084}_{-0.084}$ \\
        \hline \hline
      \end{tabular}
  \end{center}
  \vspace{-3ex}
  \caption{\label{tab:r3c}Predicted ratios $R^3$ at the LO
  from various PDFs with the PDF+$\alpha_s$ uncertainties at 68\% C.L.
  for model C.}
\end{table}

\section{Experimental Implications\label{sec:exp}}

\subsection{Total cross section measurement at the Tevatron and LHC\label{sec:exp1}}

The ratios $R^1$, especially the ratios of the total cross sections
from the LHC to Tevatron, show a large distinction between gluon or
heavy-quark initiated cases (model A or B) and the light-quark case
(model C). For example, the central predictions for $R^1_{L7/T}$ are
17.1, 24.2, and 4.0 for the three models respectively according to
Tables.~\ref{tab:r1a}-\ref{tab:r1c}. With the full data sample, the
combined Tevatron measurements of the inclusive cross sections of
the new resonance are summarized in Ref.~\cite{Aaltonen:2013kxa}.
Corresponding measurements from the LHC at 7 and 8 TeV can be found
in~\cite{Aad:2012tfa,Chatrchyan:2012ufa}. We show all the measured
cross sections from different decay channels in
Table.~\ref{tab:back}, which are normalized to the predictions of
the SM Higgs boson. Note that for the $\tau\tau$ channel we show the
recent updated results instead~\cite{atau,ctau}. The ATLAS and CMS
results are combined here by taking a weighted average with weights
of one over square of the corresponding experimental errors. Thus
correlations of systematic uncertainties in the two experiments are
simply neglected, resulting in optimistic estimations of the
combined uncertainties. Most of the results shown are for the
inclusive productions, which also receive contributions from the
Higgs-strahlung or VBF final states. Presumably they are only a
small fraction compared to the ones from the direct production in
the experimental analyses. It can be seen that the experimental
errors, especially the ones from the Tevatron are far above the
theoretical ones shown in Tables.~\ref{tab:r1a}-\ref{tab:r1c}. Thus
from Tables.~\ref{tab:r1a}-\ref{tab:r1c} and neglecting the
theoretical errors, we obtain the theoretical predictions for
$R^1_{L7(8)/T}$ as 1(1), 1.42(1.44), and 0.23(0.22) for models A, B,
and C respectively, using the relative strength (all cross sections
normalized to the corresponding predictions of the SM Higgs boson).
Without knowing the precise probability distribution of the
experimental measurements we simply assume they are Gaussian
distributed with the errors symmetrized. Based on the two data
points ($\gamma\gamma$ and $WW^*$ channels) we calculate the
$\chi^2$ values as 1.9, 2.4 and 3.3 for the models A, B, and C,
respectively. Thus all three models agree well with the current
data. The predictive power of $R^1_{L7(8)/T}$ is mostly limited by
the large experimental errors from Tevatron. However, further
precise measurements from the LHC may show improvements on
discriminations of the three models. For example, assuming the
central measurements to be exactly the same as the SM predictions
and the fractional errors reduced to 20\% for both the
$\gamma\gamma$ and $WW^*$ channels, the $\chi^2$ for model C would
be 8.4, corresponding to an exclusion at 98.5\% C.L.

We can also look at
the ratios $R^1$ at the LHC with different energies. But they are not so distinguishable
among different initial states since the light quarks there are mostly sea-like for the corresponding
energies. For the model C, using the relative strength the predictions for
$R^1_{L14/L7(8)}$ are 0.70(0.76), which require a high experimental precision in order to
distinguish them with the SM predictions with values 1(1).

\begin{table}[h!]
  \begin{center}
      \begin{tabular}{c|ccccc}
        \hline \hline
      & $\gamma\gamma$ & $ZZ^*$ & $WW^*$ & $\tau\tau$ & combined \\
    \hline
    Tevatron & $6.0^{+3.4}_{-3.1}$ & -- & $0.94^{+0.85}_{-0.83}$ & -- &-- \\
    \hline
    ATLAS & $1.8^{+0.5}_{-0.5}$ & $1.2^{+0.6}_{-0.6}$  & $1.3^{+0.5}_{-0.5}$  & $1.4^{+0.5}_{-0.4}$  & $1.4^{+0.3}_{-0.3}$\\
    \hline
    CMS & $1.4^{+0.6}_{-0.6}$ & $0.7^{+0.5}_{-0.4}$ & $0.7^{+0.5}_{-0.5}$ & $1.1^{+0.4}_{-0.4}$ & $0.87^{+0.23}_{-0.23}$ \\
    \hline
    ATLAS+CMS & $1.6^{+0.4}_{-0.4}$ & $0.9^{+0.4}_{-0.4} $ & $1.0^{+0.4}_{-0.4}$ & $1.2^{+0.3}_{-0.3}$ & $1.1^{+0.2}_{-0.2}$ \\
    \hline \hline
      \end{tabular}
  \end{center}
  \vspace{-3ex}
  \caption{\label{tab:back}Measured production cross sections of the new resonance through
  different decay channels at the Tevatron and LHC (7 and 8 TeV combined). All values are
  normalized to the corresponding cross sections of the SM Higgs boson production. The ATLAS and CMS
  results are combined by taking a weighted average neglecting correlations.}
\end{table}

\subsection{Expectations from the centrality ratios\label{sec:exp2}}

The centrality ratios $R^2$ at the LHC also display moderate
differences between the model A or B and the model C. To measure the
rapidity of the resonance we need to fully reconstruct the final
state kinematics. Thus the most promising decay channels for
measuring $R^2$ are $\gamma\gamma$ and $ZZ^*$. As shown in
Tables.~\ref{tab:r2a}-\ref{tab:r2c}, the theoretical errors for the
predictions of $R^2$ are a few percents and are rather small
compared to the experimental ones. The central predictions for
$R^2_{L14}$ are 0.45 and 0.31 for the SM and model C. For both of
the two decay channels the experimental errors of $R^2$ are expected
to be dominated by the statistical errors whether due to the low
event rate or large backgrounds. At the LHC 7 TeV (5.1 $fb^{-1}$),
for the diphoton channel after all the selection cuts, the CMS
measurement expects about 77 signal events and 311 events per GeV
(invariant mass window) from the backgrounds for the case of the SM
Higgs boson~\cite{Chatrchyan:2012ufa}. If we assume a 100 $fb^{-1}$
data sample at 14 TeV from each of the CMS and ATLAS experiments,
and assume the same event selection efficiencies, the expected event
numbers within a mass window of 4 GeV will be about $1.0\times 10^4$
for the SM Higgs boson and $1.1\times 10^5$ for the
backgrounds.\footnote{We use $R^1_{L14/L7}$ from the model C to
convert the background rate from 7 TeV to 14 TeV since they are both
$q\bar q$ initial state dominant.} Then the expected measurement of
$R^2_{L14}$ is about $0.45\pm 0.024$ including only the statistical
error.\footnote{As an estimation we simply assume the backgrounds
have the same rapidity profile as the signal of the model C for the
calculation of the statistical errors.} Thus for this case we may
exclude the model C (with $R^2_{L14}$=0.31) at $5\sigma$ C.L.. The
$ZZ^*\rightarrow 4l$ channel is almost background free and the
observed event number at the CMS is 9 for 7 and 8 TeV
combined~\cite{Chatrchyan:2012ufa}. With the same assumptions as the
$\gamma\gamma$ channel, the expected event rate is about 513, and
the measurement of $R^2_{L14}$ is $0.45\pm 0.036$ for $ZZ^*$
channel. The statistical error is larger than the one of the
diphoton channel but the measurement is free of the systematic
errors from the background estimations. A more comprehensive study
on $R^2$ should be done by the experimentalist to further examine
the backgrounds and all the systematic errors, which may change the
conclusions here.

\subsection{Observability of the associated production\label{sec:exp3}}

\begin{figure}[h]
  \begin{center}
    \includegraphics[width=0.38\textwidth]{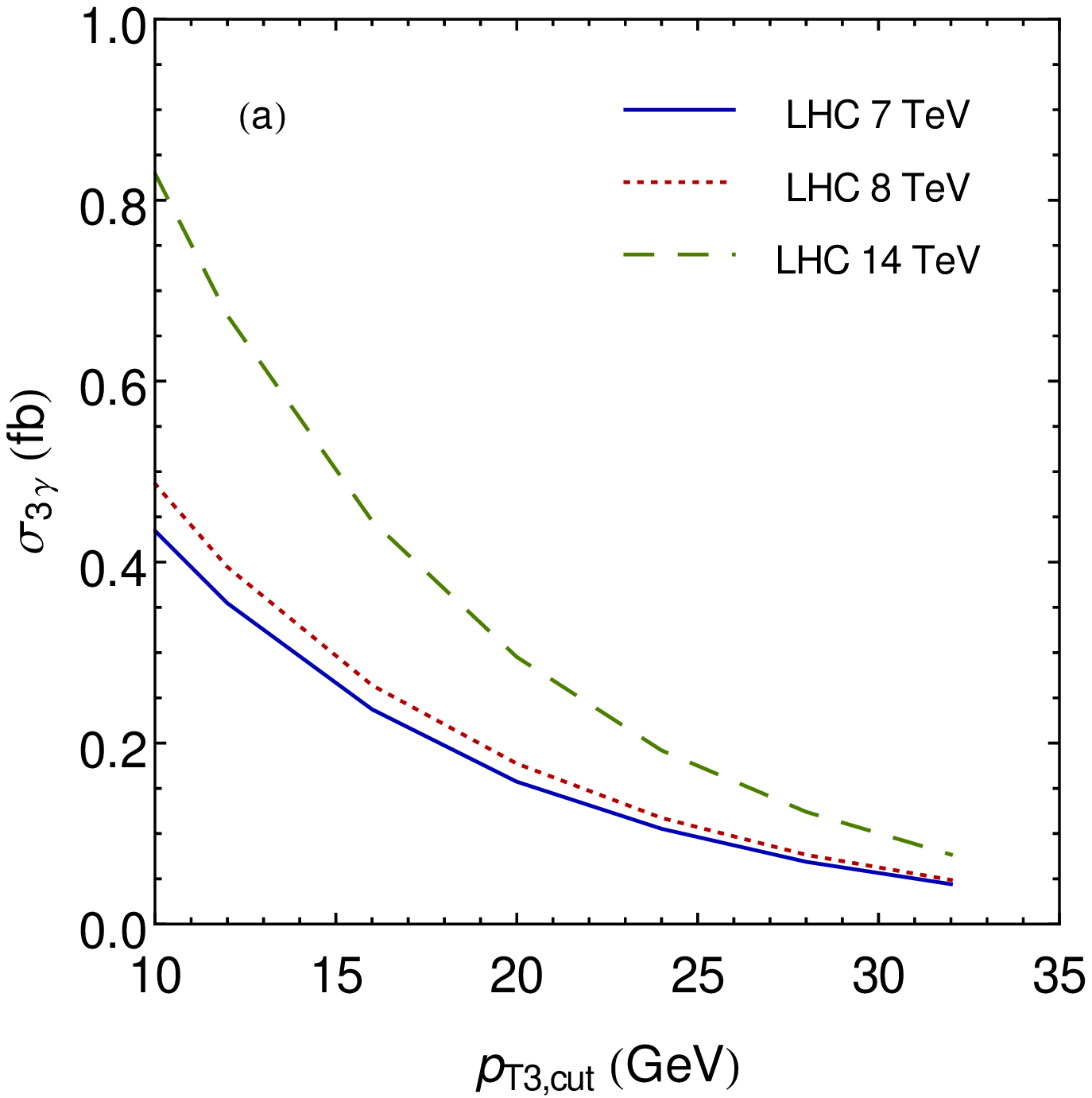}
    \includegraphics[width=0.4\textwidth]{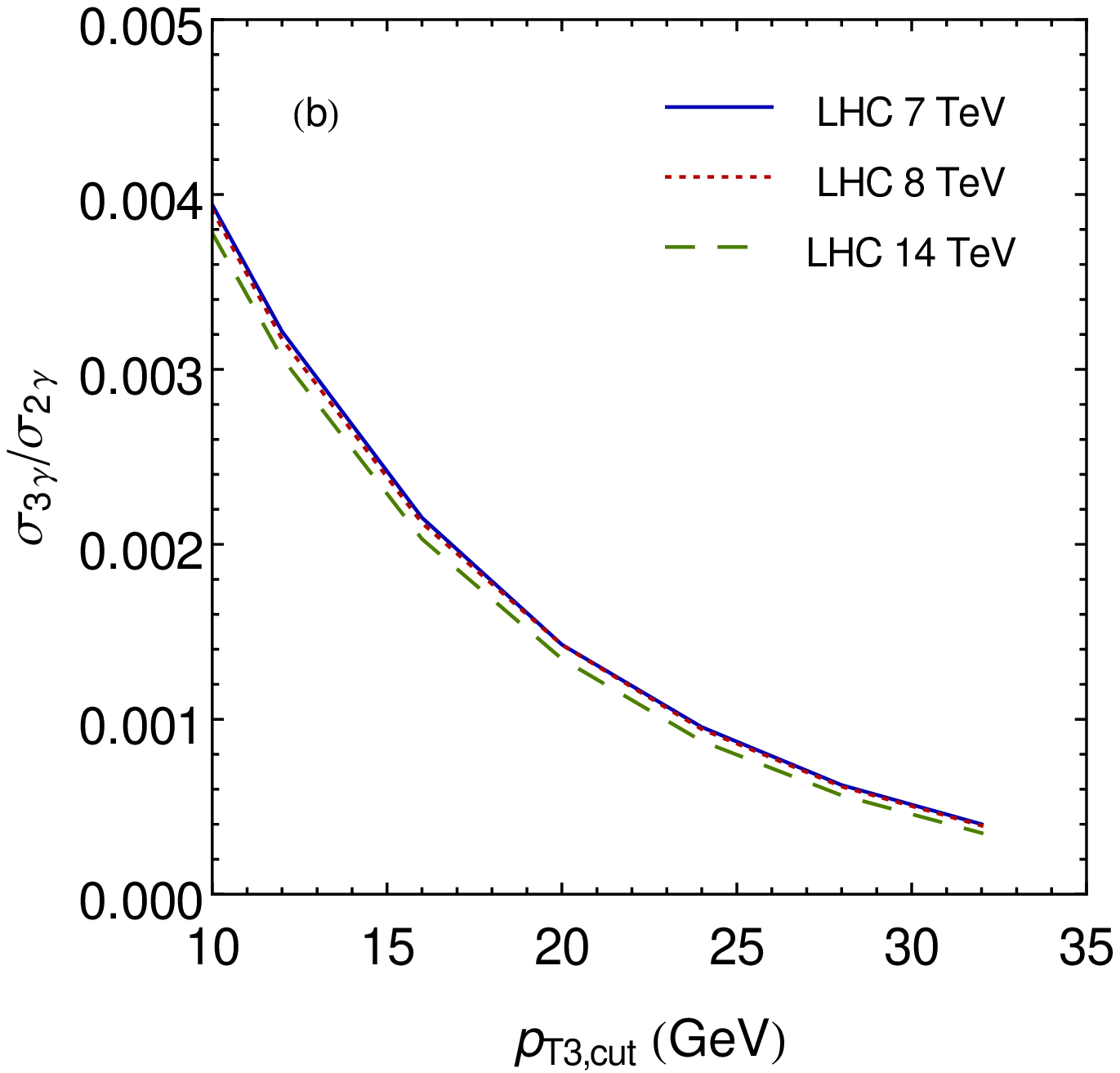}
  \end{center}
  \vspace{-5ex}
  \caption{\label{fig:back}
  (a), Cross sections of the SM tri-photon production at the LHC; (b), ratios
  of the cross sections of the SM tri-photon production to the ones of diphoton production.
  The selection cuts are applied to both the tri-photon and diphoton events.}
\end{figure}

The associated production of the SM Higgs boson with photon is
almost unobservable at the LHC with a rate of less than $10^{-4}$ of
the direct production rate. While for models B
and C the rates are an order of magnitude higher. Even though they
may be still difficult to be observed. In order to suppress the
backgrounds and obtain sufficient statistics, the diphoton decay
channel is the only realistic solution. Thus we need to look at the
tri-photon final state. As a quick estimation for the background, we
can calculate the ratios of the cross sections of the SM direct
tri-photon production (intrinsic backgrounds) to the ones of
diphoton production. The selection cuts for the two or three photon
events ($p_T$ ordered) are as below
\begin{eqnarray}
&&|\eta_{\gamma}|<2,\, \Delta R_{\gamma\gamma}>0.4,\, p_{T,1}>30\,{\rm GeV},\, p_{T,2}>20\,{\rm GeV},\,\nonumber\\
&&p_{T,3}>15\,{\rm GeV},\, 124<m_{12}<128\,{\rm GeV}.
\end{eqnarray}
Here both the cross sections of the diphoton and tri-photon
productions are calculated at the LO using Madgraph
4~\cite{Alwall:2007st}. Contributions from quark fragmentations and
gluon-initiated loop diagrams are not included. We plot the cross
section $\sigma_{3\gamma}$ as well as the ratio
$\sigma_{3\gamma}/\sigma_{2\gamma}$ as functions of the $p_T$
threshold of the softest photon in the tri-photon production in
Fig.~\ref{fig:back}. The ratios are similar to the results of the
models B and C shown in Fig.~\ref{fig:r3}, with
$\sigma_{3\gamma}/\sigma_{2\gamma} \sim 0.0022$ for
$p_{T,3}>15\,{\rm GeV}$.

By the same assumptions as in Section~\ref{sec:exp2}, the expected
background event rate is about $1.1\times 10^5$ for the diphoton
channel at the LHC of 14 TeV and $\mathcal{L}=100\, fb^{-1}$. Simply
multiplying it with the ratio $\sigma_{3\gamma}/\sigma_{2\gamma}$,
we estimate a background rate of about 242 for the tri-photon final
state. Similarly, using the numbers in
Tables.~\ref{tab:r3a}-\ref{tab:r3c}, the expected signal event rates
are about $0.8$, $16$ and $24$ for the SM, models B and C
respectively. We can see that the signal rates of the models B and C
are of the similar size as the $1\sigma$ statistical fluctuation of the
background. Thus although the associated production mode shows a
large distinction between the SM and the alternative models, i.e.,
models B and C, but it requires a high luminosity for the
experimental measurements, e.g., around 900 (400) $fb^{-1}$ in order to
discriminate the SM with the model B (C) at $3\sigma$ C.L. Also
note that for a variation of the model B where the charm quark coupling
is dominant instead of the bottom quark, the associated production rate
can be further enhanced by about a factor of 4 from the electric charge.

\section{Conclusions\label{sec:con}}
We performed a study on differentiating the direct production
mechanisms of the newly discovered Higgs-like boson at the LHC based
on several inclusive observables introduced, including the ratios of
the production rates at different colliders and energies, the
centrality ratios of the resonance, and the ratios of the rates of
associated production with a photon to the ones of direct production.
Above ratios reveal neither the parton constituents nor initial
state radiations involved in the production mechanisms, and are
independent of the couplings to the decay products. We select three benchmark
models, including the SM Higgs boson, to illustrate how the
theoretical predictions of the above ratios are different for the
$gg$, $b\bar b(c\bar c)$, and $q\bar q$ (flavor universal) initial
states in the direct production. The theoretical uncertainties of
the predictions are also discussed. All three models are found to
be in good agreement with inclusive rate ratios from current measurements at
the Tevatron and LHC.
Moreover, we show
expectations from further LHC measurements with high luminosities.
The centrality ratio measurements are supposed to be able to separate the
$gg$ or $b\bar b(c\bar c)$ initial states with $q\bar q$.
The tri-photon signal from the associated production may even
differentiate the $gg$ initial states with $b\bar b(c\bar c)$ or
$q\bar q$ in the direct production.

\begin{acknowledgments}
This work was supported by the U.S. DOE Early Career Research Award
DE-SC0003870 by Lightner-Sams Foundation. We appreciate insightful discussions
with Pavel Nadolsky, Stephen Sekula, Ryszard Stroynowski, and C.-P. Yuan.
\end{acknowledgments}

\end{document}